\documentclass[%
aps, %
prl, 
twocolumn, 10pt, 
superscriptaddress, %
a4paper, %
longbibliography,
amsfonts, amsmath, amssymb]{revtex4-2}
\usepackage[sort&compress]{natbib}
\usepackage{newtxtext}
\usepackage[varg]{newtxmath}
\usepackage{ascmac}
\usepackage{mathrsfs}
\usepackage{bm}
\usepackage{color}
\usepackage{graphicx}
\usepackage{bm}
\usepackage{braket}
\usepackage{mathrsfs}
\usepackage[normalem]{ulem}
\usepackage{hhline}
\usepackage{hyperref}
\hypersetup{%
    unicode=false, %
    bookmarksnumbered=true, %
    bookmarkstype=toc, %
    breaklinks=true, %
    colorlinks=true, %
    pdfborder={0 0 1}, %
    linkcolor=blue, %
    citecolor=blue, %
    urlcolor=blue, %
    pdftoolbar=true, %
    pdfmenubar=true, %
    pdffitwindow=false, %
    pdfstartview={FitH} %
}
\begin{document}
%
%
%
%
%
\title{Electron Hydrodynamics by Spin Hall Effect}
\date{\today}
\author{Junji Fujimoto}
\affiliation{Department of Electrical Engineering, Electronics, and Applied Physics, Saitama University, Saitama, 338-8570, Japan}
\email[E-mail address: ]{jfujimoto@mail.saitama-u.ac.jp}

\author{Wataru Koshibae}
\affiliation{RIKEN Center for Emergent Matter Science (CEMS), Wako, Saitama 351-0198, Japan}

\author{Sadamichi Maekawa}
\affiliation{RIKEN Center for Emergent Matter Science (CEMS), Wako, Saitama 351-0198, Japan}
\affiliation{Kavli Institute for Theoretical Sciences, University of Chinese Academy of Sciences, Beijing, 100190, China}
\affiliation{Advanced Science Research Center, Japan Atomic Energy Agency, Tokai 319-1195, Japan}

\begin{abstract}
Electron hydrodynamics is currently known to emerge only when electron-electron interaction dominates over the momentum-nonconserving scatterings of electrons, where the electron transport is described by a hydrodynamic equation.
Here we show that electron transport in electron systems with the spin Hall effect is also given by the hydrodynamic equation, whose kinetic viscosity is determined by the spin diffusion length and the transport lifetime.
The electric current vorticity is proportional to the spin accumulation due to the spin Hall effect in two-dimensional systems.
We demonstrate by solving the hydrodynamic equation in a two-dimensional system with a cavity, combined with micromagnetic simulation for an attached chiral magnetic insulator, that the spin accumulated near the boundary of the cavity creates a magnetic skyrmion.
Our findings and demonstration shed light on a novel aspect of electron hydrodynamics and spin transport.
\end{abstract}
\maketitle
Recent experiments show that electron transport is beyond Ohm's law and becomes hydrodynamic in some materials with high purity, such as semiconductor systems~\cite{dejong1995,gupta2021,keser2021}, graphene~\cite{bandurin2016,sulpizio2019,ku2020}, $\mathrm{Pd Co O_2}$~\cite{moll2016}, and $\mathrm{W Te_2}$~\cite{vool2021}, which has arisen a research field of electron hydrodynamics.
The electron transport in the hydrodynamic regime is described by the Stokes equation with the frictional force~\cite{torre2015,polini2020}.
One typical feature of the hydrodynamic flow is nonzero electric current vorticity $\bm{\omega}_e = \bm{\nabla} \times \bm{j}_e$, where $\bm{j}_e$ is the electric current density, while the vorticity is zero for the current obeying Ohm's law in the steady state~\footnote{In the cases where the electric conductivity is varying spatially, such as in gradient materials~\cite{okano2019}, the electric current vorticity is finite even when the Ohm's law is realized, but the viscosity in such systems is zero.}.
Electron hydrodynamics is currently known to emerge only in the specific condition that the electron-electron interaction overcomes the scatterings of electrons without momentum conservation.

Electron transport in systems with the direct and inverse spin Hall effects~\cite{dyakonov1971,sinova2015} can be regarded beyond Ohm's law, $\bm{j}_e^{\mathrm{Ohm}} = \sigma_e \bm{E}$, where $\sigma_e$ is the electric conductivity of the material, and $\bm{E}$ is the applied electric field.
The inverse spin Hall effect induces an additional electric current density, $\bm{j}_{e}^{\mathrm{SH}} = \theta_{\mathrm{SH}} \sum_{\alpha} \hat{\alpha} \times \bm{j}_{\mathrm{s}}^{\alpha}$ with $\alpha = x, y, z$, where $\bm{j}_{\mathrm{s}}^{\alpha}$ is the spin current (in the unit of the electric current density) whose spin polarization is directed to $\hat{\alpha}$, $\theta_{\mathrm{SH}}$ is the spin Hall angle giving the conversion efficiency, and $\hat{\alpha}$ is the $\alpha$-directed unit vector.
Hence, the total electric current density is given by
\begin{align}
\bm{j}_e
    & = \bm{j}_e^{\mathrm{Ohm}} + \bm{j}_{e}^{\mathrm{SH}}
    = \sigma_e \bm{E} + \theta_{\mathrm{SH}} \sum_{\alpha} \hat{\alpha} \times \bm{j}_{\mathrm{s}}^{\alpha}
\label{eq:j_e}
.\end{align}
Equation~(\ref{eq:j_e}) indicates that electron transport in systems with the spin Hall effect is beyond Ohm's law due to the contribution from the inverse spin Hall effect, $\bm{j}_e^{\mathrm{SH}}$.
Note that the zero vorticity for the current obeying Ohm's law in the steady state is simply proven as, $\bm{\omega}_e^{\mathrm{Ohm}} = \bm{\nabla} \times \bm{j}_e^{\mathrm{Ohm}} = - \sigma_e \bm{\nabla} \times \bm{\nabla} \phi = 0$ with $\bm{E} = - \bm{\nabla} \phi$, where $\phi$ is the scaler potential.

In this Letter, we show that the electron transport in the systems with the spin Hall effect (we call such systems the spin Hall systems) is also described by the Stokes equation as in electron hydrodynamics.
This indicates that the electron transport in the spin Hall systems becomes hydrodynamic even for no electron-electron interactions.
The kinetic viscosity in the spin Hall systems is determined by the spin diffusion length and the transport lifetime.
In two-dimensional~(2D) spin Hall systems, the electric current vorticity is proportional to the spin accumulation due to the spin Hall effect, while the vorticity in the three-dimensional systems is partially given by the spin accumulation.
We also discuss the boundary condition between the spin Hall system and the vacuum, which is different from that of the electron hydrodynamics caused by the electron-electron interaction.
Furthermore, we propose a novel guiding principle to manipulate spin accumulation by geometrical structures and to induce topological transitions of magnetic textures based on the electron hydrodynamic viewpoint.
As a demonstration, we show by solving the Stokes equation combined with the micromagnetic simulation for a bilayer of a specific spin Hall system and chiral magnetic insulator with cavity structure, that the spin accumulation near the cavity creates a magnetic skyrmion.

We begin with Eq.~(\ref{eq:j_e}) and
\begin{align}
\bm{j}_{\mathrm{s}}^{\alpha}
    & = - \frac{\sigma_e}{e} \bm{\nabla} \mu_{\mathrm{s}}^{\alpha}
    + \theta_{\mathrm{SH}} \hat{\alpha} \times \bm{j}_e
\label{eq:j_s}
,\end{align}
where $\mu_{\mathrm{s}}^{\alpha}$ is the spin accumulation (in the energy unit) in the $\alpha$ direction with $\alpha = x, y, z$.
The first term of Eq.~(\ref{eq:j_s}) describes the diffusion spin current and the second term is caused by the direct spin Hall effect.
In phenomenology, the spin accumulation in the steady state is distributed based on the spin diffusion equation,
\begin{align}
\nabla^2 \mu_{\mathrm{s}}^{\alpha}
    & = \frac{\mu_{\mathrm{s}}^{\alpha}}{\lambda_{\mathrm{s}}^2}
\label{eq:SDE}
,\end{align}
where $\lambda_{\mathrm{s}}$ is the spin diffusion length.
We note that the spin accumulation is directly observed by the optical Kerr effect~\cite{kato2004,sih2005,stamm2017}, by the circular polarization of light~\cite{wunderlich2005}, and by X-ray~\cite{ruiz-gomez2022} and indirectly by the so-called spin Hall magnetoresistance~\cite{nakayama2013,kim2016}.
Inserting Eq.~(\ref{eq:j_s}) into Eq.~(\ref{eq:j_e}), we have 
\begin{align}
\bm{j}_e
    & = \sigma'_e \bm{E} + \theta_{\mathrm{SH}} \frac{\sigma'_e}{e} \bm{\nabla} \times \bm{\mu}_{\mathrm{s}}
\label{eq:j_e2}
,\end{align}
where $\sigma'_e$ is the renormalized electric conductivity given by
\begin{align}
\sigma'_e
    & = \frac{\sigma_e}{1 + 2 \theta_{\mathrm{SH}}^2}
\label{eq:sigma'_e}
.\end{align}
Sometimes Eq.~(\ref{eq:j_e2}) in which $\sigma'_e$ is replaced with $\sigma_e$ is used as a starting point in phenomenological treatments~\cite{chen2013}.
In the Supplemental Information~(SI)~\cite{SM}, we discuss the relation between Eqs.~(\ref{eq:j_e}) and (\ref{eq:j_e2}).
Then, taking the rotation of Eq.~(\ref{eq:j_e2}) twice, and noting that $\bm{\nabla} \times \bm{\omega}_e = - \nabla^2 \bm{j}_e$ because of the charge continuity equation in the steady state, $\bm{\nabla} \cdot \bm{j}_e = 0$, we get $\lambda_{\mathrm{s}}^2 \nabla^2 \bm{j}_e = (\theta_{\mathrm{SH}} \sigma'_e / e) \bm{\nabla} \times \bm{\mu}_{\mathrm{s}}$, where we have used Eq.~(\ref{eq:SDE}).
Using this relation, we eliminate the spin accumulation from Eq.~(\ref{eq:j_e2}), so that we obtain
\begin{align}
- \sigma'_e \bm{\nabla} \phi + \lambda_{\mathrm{s}}^2 \nabla^2 \bm{j}_e - \bm{j}_e = 0
\label{eq:Stokes}
,\end{align}
which is exactly the same as the Stokes equation in electron hydrodynamics~\cite{torre2015,polini2020}.
The diffusion length of the vorticity in the present case is given by the spin diffusion length.
Equation~(\ref{eq:Stokes}) is the first main result of this work, which indicates that the electron transport in spin Hall systems is described by the electron hydrodynamic equation.
The kinetic viscosity is given by the spin diffusion length and the transport lifetime $\tau$ as $\nu = \lambda_{\mathrm{s}}^2 / \tau$.
The steady distribution of the electric current density in the spin Hall systems is better estimated from the Stokes equation~(\ref{eq:Stokes}) with $\bm{\nabla} \cdot \bm{j}_e = 0$, not from Ohm's law with the Laplace equation.

Next, we show that the electric current vorticity is directly connected to the spin accumulation in the 2D spin Hall systems.
We suppose that $\bm{j}_e$ flows in the $xy$-plane, so that the vorticity is in the $z$ direction as well as the spin accumulation due to the spin Hall effect $\bm{\mu}_{\mathrm{s}} = \mu_{\mathrm{s}}^z \hat{z}$.
By taking the rotation of Eq.~(\ref{eq:j_e2}) and using Eq.~(\ref{eq:SDE}), we obtain
\begin{align}
\bm{\omega}_e^{\mathrm{2D}}
    & = - \frac{\theta_{\mathrm{SH}} \sigma'_e}{e \lambda_{\mathrm{s}}^2} \mu_{\mathrm{s}}^z \hat{z}
\label{eq:omega_2D}
,\end{align}
which shows that the electric current vorticity is directly connected to the spin accumulation.
The hydrodynamic nature of the spin Hall systems is represented by the spin accumulation.
The spin accumulation at the boundaries is generated by the direct spin Hall effect, which is proportional to the spin Hall angle.
The generated spin accumulation affects again the charge transport as the vorticity via the inverse spin Hall effect.
Hence, the vorticity is proportional to $\theta_{\mathrm{SH}}^2$, whereas the spin accumulation is of the order of $\theta_{\mathrm{SH}}$, and the spin accumulation and the vorticity are connected via Eq.~(\ref{eq:omega_2D}).
The point is that the charge and spin transports in the spin Hall systems are connected to each other since the direct and the inverse spin Hall effects arise simultaneously.
We emphasize that the spin accumulation is equivalent to the electric current vorticity in the 2D spin Hall systems, which is not a causal relation between the spin accumulation and the vorticity, but is correspondence relation originating from the spin-orbit coupling.
This correspondence allows us to access the spin transport in the 2D spin Hall systems by solving the Stokes equation and taking the rotation of the electric current density.
Equation~(\ref{eq:Stokes}) together with Eq.~(\ref{eq:omega_2D}) provides the self-consistent treatment of charge and spin transports in the spin Hall systems with complex geometrical structures, based on the sophisticated technique of hydrodynamics.

In the three-dimensional cases, the vorticity partially captures the spin accumulation; $(\theta_{\mathrm{SH}} \sigma'_e / e) \bm{\mu}_{\mathrm{s}} = - \bm{\nabla} \psi_{\mathrm{s}} + \lambda_{\mathrm{s}}^2 \bm{\omega}_e^{\mathrm{3D}}$, where $\psi_{\mathrm{s}}$ is the rotationless component of the spin accumulation when the spin accumulation is decomposed based on the Helmholtz decomposition.
Note that the Stokes equation~(\ref{eq:Stokes}) is realized also for the three-dimensional spin Hall systems.

The behavior near a boundary of the hydrodynamic flow in the spin Hall systems is different from that of the electron hydrodynamics caused by the electron-electron interaction, for the boundary between the electron systems and the vacuum.
The conventional electron hydrodynamic flow decreases near the boundary, based on the physical insight of viscosity.
The reduction near the boundary is sometimes expressed by the slip length~\cite{kiselev2019}.
The finite slip length indicates the corresponding decrease of the electric current density, and the infinite slip length means zero reduction.
More importantly, the electric current density is considered to decrease near the boundary and the slip length would not be negative.
However, the electric current density \textit{increases} as going near the boundary in the spin Hall systems.
The boundary condition originates from the fact that the spin current across the boundary should be zero; $\bm{j}_{\mathrm{s}}^{\alpha} |_{\mathrm{boundary}} = 0$ yields $(\sigma_e / e) \bm{\nabla} \times \bm{\mu}_{\mathrm{s}} |_{\mathrm{boundary}} = 2 \theta_{\mathrm{SH}} \bm{j}_e |_{\mathrm{boundary}}$ from Eq.~(\ref{eq:j_s}), and hence we obtain $\bm{j}_e |_{\mathrm{boundary}} = \sigma_e \bm{E} |_{\mathrm{boundary}}$ by inserting it into Eq.~(\ref{eq:j_e2}).
Since the bare electric conductivity is larger than the renormalized one; $\sigma_e > \sigma'_e$ [see Eq.~(\ref{eq:sigma'_e})], we conclude that the electric current density increases as getting close to the boundary in the spin Hall systems.
Its physical meaning is simple; the spin accumulation is induced by the direct spin Hall effect near the boundary, and the diffusion spin current of the accumulation is converted to the additional electric current due to the inverse spin Hall effect, so that the electric current density increases near the boundary.

\begin{figure}[tb]
\centering
\includegraphics[width=\linewidth]{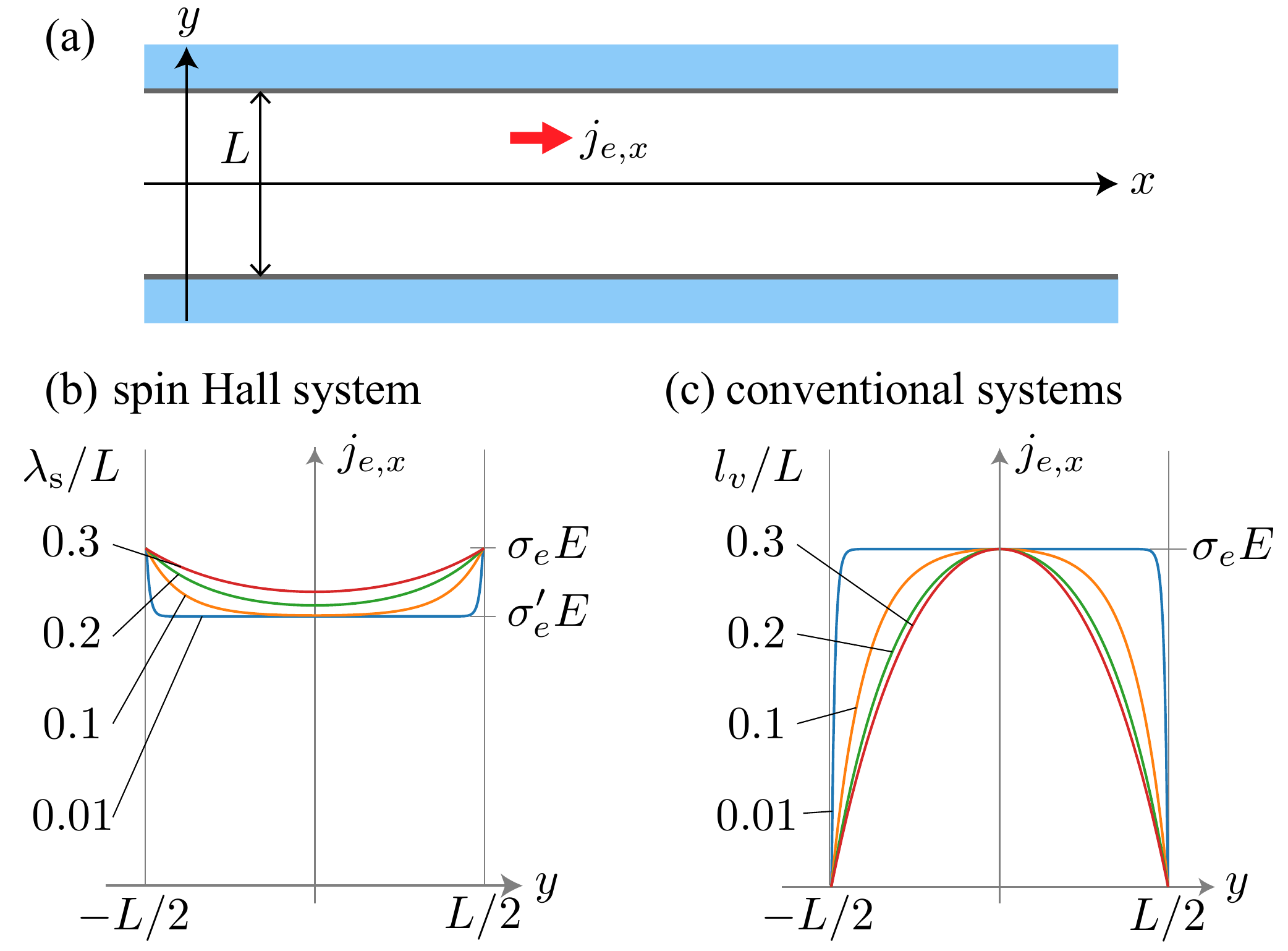}
\caption{\label{fig:BC}Typical example of a 2D pipe flow for the spin Hall and conventional electron hydrodynamic systems.
(a)~Geometrical setup of the pipe flow. The pipe width is given by $L$, the outside of the pipe is supposed to be in vacuum.
The hydrodynamic viscous flow along the pipe-length ($x$-) direction has a spatial distribution depending on the pipe-width ($y$-) direction due to the viscosity.
(b)~Electric current density distribution in the spin Hall system for various spin diffusion length $\lambda_{\mathrm{s}}$. The electric current density increases near the boundary.
(c)~Electric current density distribution in conventional electron hydrodynamic systems.
The boundary condition is set to be $j_{e, x} (y = \pm L / 2) = 0$ corresponding to the no-slip condition.
}
\end{figure}
To understand the difference between the behaviors near the boundary of the spin Hall systems and the conventional electron hydrodynamic flow, we consider a typical example of the 2D pipe flow~[Fig.~\ref{fig:BC}~(a)].
Solving the Stokes equation~(\ref{eq:Stokes}) for the geometrical configuration of 2D pipe flow, the general solution is given by $j_{e, x} (y) = A e^{y / \lambda_{\mathrm{s}}} + B e^{- y / \lambda_{\mathrm{s}}} + \sigma'_e E_x$, where $E_x = - \partial \phi / \partial x = \mathrm{const}.$ is the applied electric field, and $A$ and $B$ are the unknown parameters that should be determined by the boundary conditions.
The boundary conditions are given by $j_{e, x} (y = \pm L / 2) = \sigma_e E_x$, so that we obtain $j_{e, x} (y) = [1 + 2 \theta_{\mathrm{SH}}^2 \cosh (y / \lambda_{\mathrm{s}}) / \cosh (L / 2 \lambda_{\mathrm{s}})] \sigma'_e E_x$.
Figure~\ref{fig:BC}~(b) shows the electric current density distribution for the 2D spin Hall system.
Meanwhile, the Stokes equation~(\ref{eq:Stokes}) for the conventional electron hydrodynamics, which is obtained by replacing $\sigma'_e$ and $\lambda_{\mathrm{s}}^2$ with $\sigma_e$ and $\nu \tau$, is also solved as $j_{e, x} (y) = \sigma_e E_x [\cosh (y/l_v) - \cosh (L / 2 l_v)] / [1 - \cosh (L / 2 l_v)]$, where $l_v = \sqrt{\nu \tau}$ is the diffusion length of the vorticity~(see SI for the derivation).
Although there is an ambiguity in choosing the boundary condition regarding the slip length, we have set $j_{e, x} (y = \pm L / 2) = 0$ (the no-slip condition).
Figure~\ref{fig:BC}~(c) depicts the electric current density distribution for the various diffusion lengths of vorticity $l_v$ in the case of the conventional electron hydrodynamic flow.
Figures~\ref{fig:BC} (b) and (c) reveal the difference between the spin Hall system and the conventional hydrodynamic flow.

\begin{figure}[tb]
\centering
\includegraphics[width=\linewidth]{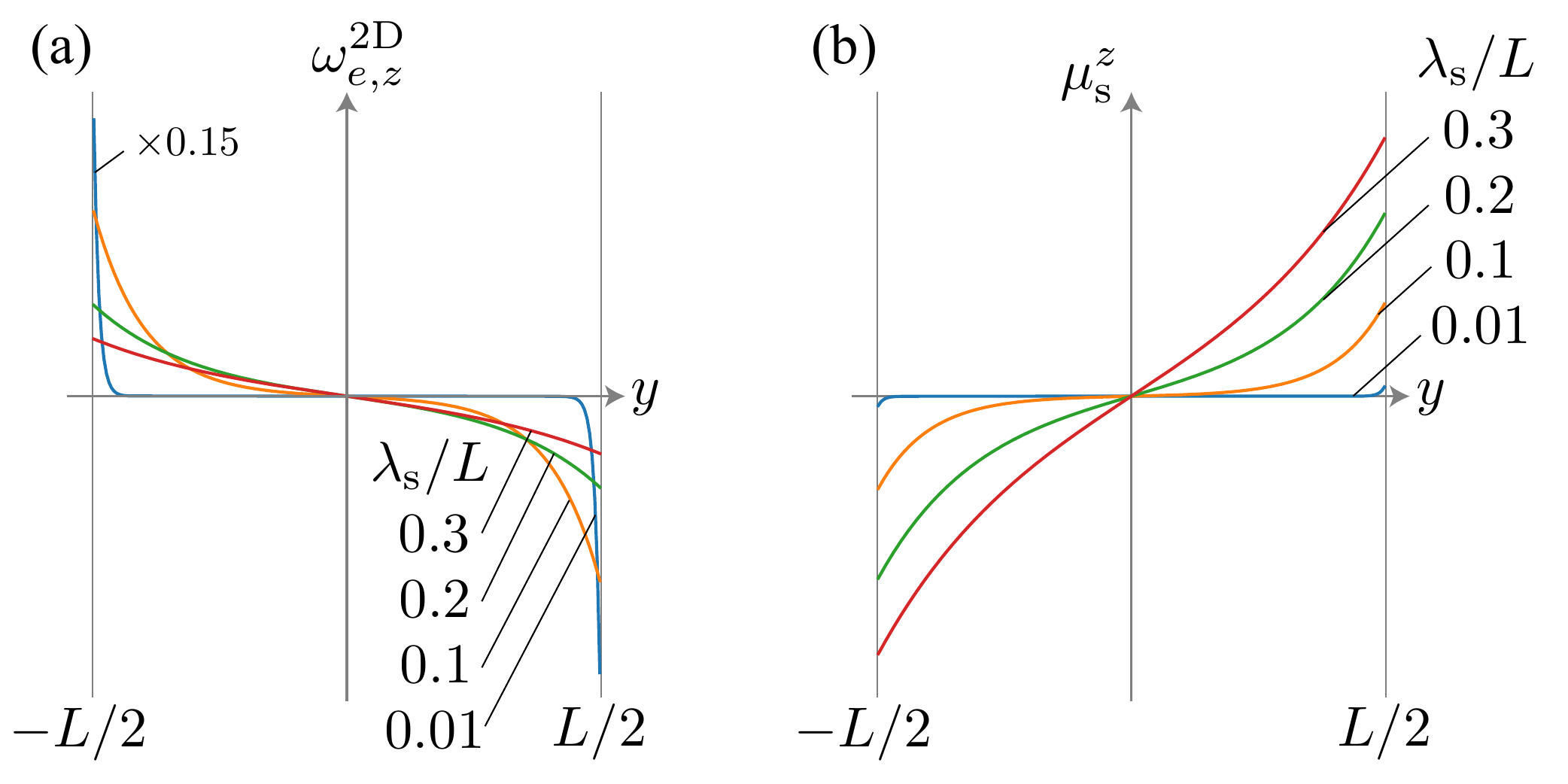}
\caption{\label{fig:w-mu}Electric current vorticity and corresponding spin accumulation of the 2D pipe flow for various spin diffusion lengths.
The vorticity and spin accumulation are connected via Eq.~(\ref{eq:omega_2D}).
The vorticity is larger near the boundary as the spin diffusion length is shorter, while the spin accumulation is smaller (see also SI).
}
\end{figure}
Here, we show that we can discuss the spin transport based on the electric current vorticity through Eq.~(\ref{eq:omega_2D}).
For the 2D pipe flow in the spin Hall system, the vorticity is obtained as $\omega_{e, z}^{\mathrm{pipe}} = - (2 \theta_{\mathrm{SH}}^2 \sigma'_e E_x / \lambda_{\mathrm{s}}) \sinh (y/\lambda_{\mathrm{s}}) / \cosh (L/2 \lambda_{\mathrm{s}})$, which is shown in Fig.~\ref{fig:w-mu}~(a).
Inserting this expression into Eq.~(\ref{eq:omega_2D}), the corresponding spin accumulation is given as $\mu_{\mathrm{s}}^{\mathrm{pipe}, z} = 2 e \lambda_{\mathrm{s}} \theta_{\mathrm{SH}} E_x \sinh (y/\lambda_{\mathrm{s}}) / \cosh (L/2 \lambda_{\mathrm{s}})$, which is plotted in Fig.~\ref{fig:w-mu}~(b).
An important point is that the vorticity is proportional to $\theta_{\mathrm{SH}}^2$, and that the spin accumulation is proportional to $\theta_{\mathrm{SH}}$.
The spin accumulation is induced by the direct spin Hall effect and hence depends linearly on the spin Hall angle, while the electric current vorticity is arisen from the direct and inverse spin Hall effects, which leads to the dependence of $\theta_{\mathrm{SH}}^2$.
As in Fig.~\ref{fig:w-mu}~(b), the spin accumulation takes the smaller value as the spin diffusion length becomes shorter for the same spin Hall angle, which is reasonable since the shorter spin diffusion length leads to more rapid decay of the spin accumulation.

\begin{figure}[tb]
\centering
\includegraphics[width=\linewidth]{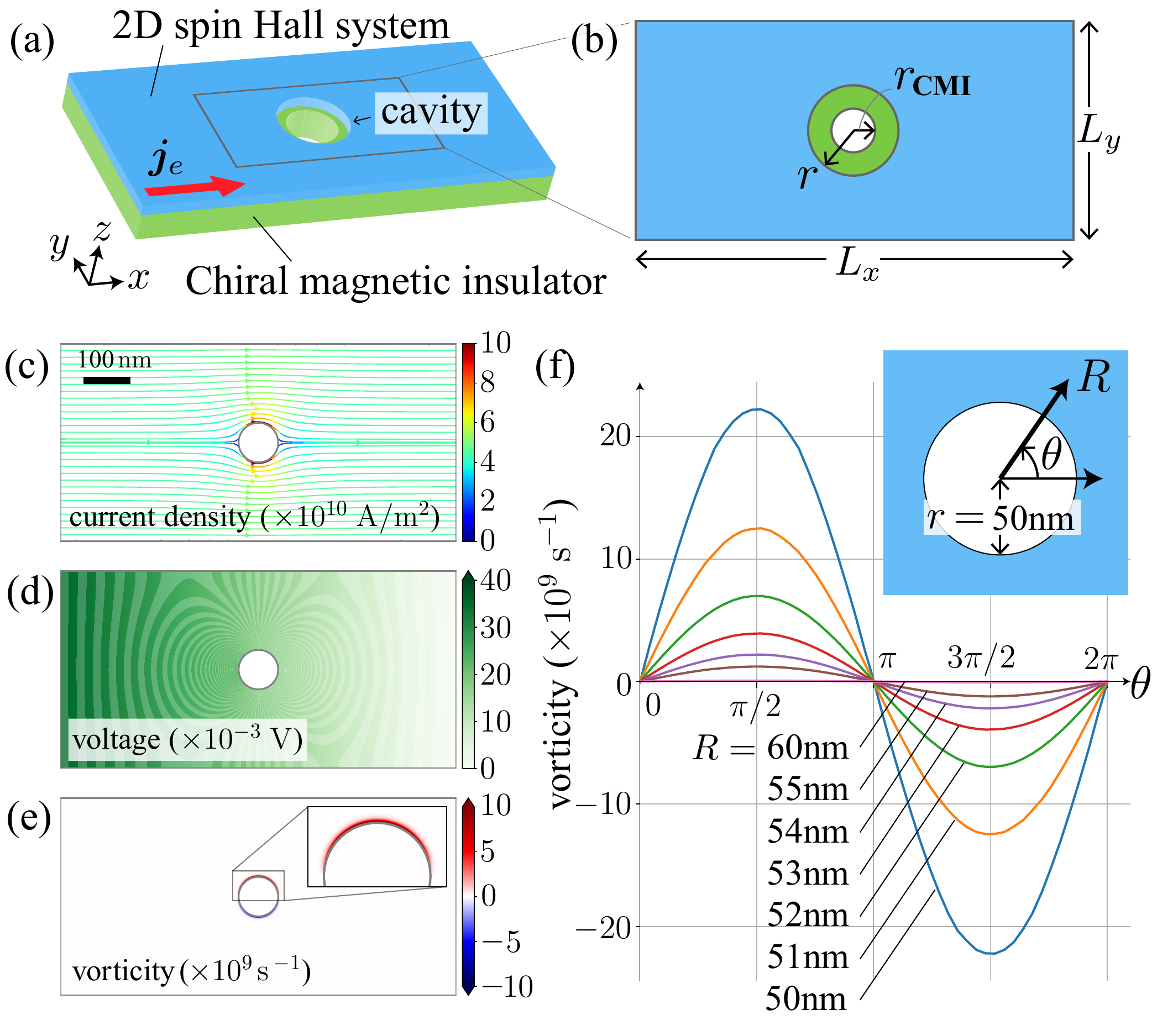}
\caption{\label{fig:FEM}Numerical simulations of the Stokes equation.
(a)~Schematics of the simulated system of the bilayer composed of 2D spin Hall system and chiral magnetic insulator~(CMI) with cavity structure.
(b)~Geometrical configuration of the simulations.
The radius of the cavity in the spin Hall system is larger than that of CMI.
(c)--(e)~Resultant profiles of (c)~the electric current density, (d)~the voltage, and (e)~the vorticity, solved based on the finite element method~(FEM).
(f)~Angular dependence of the electric current vorticity near the cavity for various distances from the center of the cavity.
The magnitude of the vorticity decays rapidly as getting far from the boundary of the cavity.
The inset of (f) depicts geometric definition of the distance $R$ and the angle $\theta$.
}
\end{figure}
Once given the hydrodynamic aspect of spin transport, the wealth of established knowledge of hydrodynamics brings a novel perspective on controlling spin accumulation.
For example, hydrodynamics commonly treats systems with geometrical structures, such as a cavity and a notch, and the flow near the structures is accompanied by nonzero vorticity for a viscous fluid.
Since the vorticity is connected to spin accumulation in the spin Hall system, we propose a guiding principle of manipulating the spin accumulation by geometrical structures.
As a demonstration to utilize the designed spin accumulation, we numerically show a magnetic skyrmion creation via the electric current vorticity by solving the Stokes equation~(\ref{eq:Stokes}) with a cavity in a specific 2D spin Hall system combined with the micromagnetic simulation on a chiral magnetic insulator attached to the spin Hall system.

As shown in Fig.~\ref{fig:FEM}~(a), we consider a bilayer system composed of a 2D spin Hall system and a chiral magnetic insulator~(CMI) with cavity structure.
We assume that the radius of the cavity in the CMI given by $r_{\mathrm{CMI}}$ is smaller than that in the spin Hall system denoted by $r$; $r > r_{\mathrm{CMI}}$, as depicted by Fig.~\ref{fig:FEM}~(b).
We solve the Stokes equation~(\ref{eq:Stokes}) combined with the charge conservation $\bm{\nabla} \cdot \bm{j}_e = 0$ with the boundary condition at cavity's boundary.
The material parameters of $\mathrm{Pt}_{0.6} (\mathrm{MgO})_{0.4}$~\cite{zhu2019} are used as the spin Hall system; the resistivity is $\rho_e = 1 / \sigma'_e = 74~\mathrm{\mu \Omega \cdot cm}$, the spin Hall angle is $\theta_{\mathrm{SH}} \sim 0.73$, and the spin diffusion length is $\lambda_{\mathrm{s}} = 1.75~\mathrm{nm}$.
See SI~\cite{SM} for more information on the FEM calculation.
Note that the Stokes equation~(\ref{eq:Stokes}) with $\bm{\nabla} \cdot \bm{j}_e = 0$ determines $\bm{j}_e$ and $\phi$ self-consistently, and the solution of $\phi$ is consistent with the Laplace equation $\nabla^2 \phi = 0$.

\begin{figure}[tb]
\centering
\includegraphics[width=\linewidth]{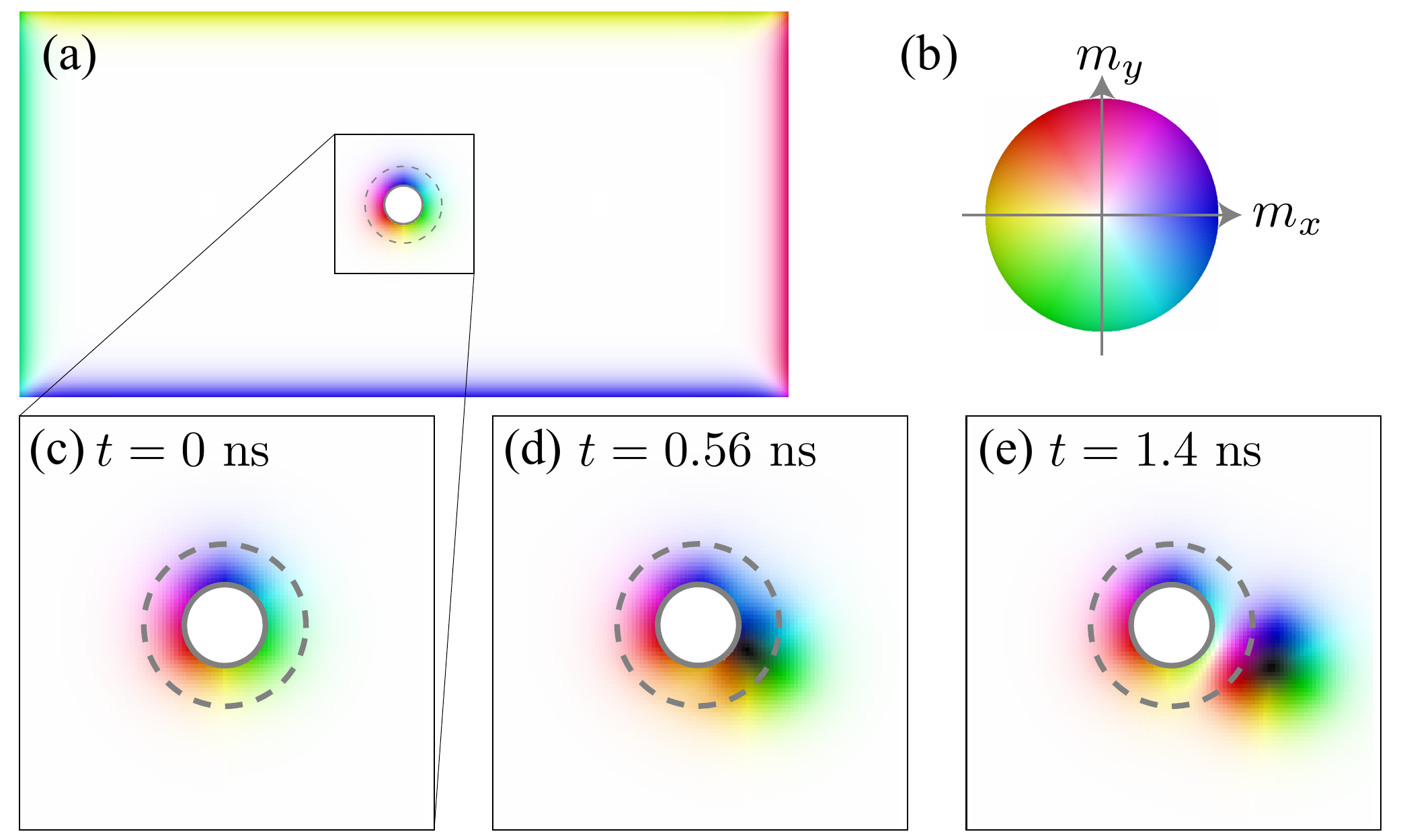}
\caption{\label{fig:skyrmion}Micromagnetic simulation result.
(a)~Snapshot of the system at $t = 0~\mathrm{ns}$.
(b)~Colormap of the magnetization.
The color code specifies the in-plane component of the magnetic moment, e.g., the blue (yellow) means the positive (negative) $m_x$ direction.
In the representation of the magnetic texture, the brightness represents the out-of-plane magnetic moment, that is, the white (black) is for $m_z = + 1$~$(- 1)$.
(c)--(e)~Snapshots near the cavity at (c)~$t = 0~\mathrm{ns}$, (d)~$t = 0.56~\mathrm{ns}$, and (e)~$t = 1.4~\mathrm{ns}$ by using the FEM result of the vorticity as the input.
The successful magnetic skyrmion creation is observed.
We set $r = 50~\mathrm{nm}$ and $r_{\mathrm{CMI}} = 25~\mathrm{nm}$.}
\end{figure}
The FEM results are shown in Fig.~\ref{fig:FEM}~(c)--(e).
Figure~\ref{fig:FEM}~(c) depicts the flow profile of the electric current density $\bm{j}_e$, and Fig.~\ref{fig:FEM}~(d) shows the spatial profile of the scaler potential $\phi$.
The scaler potential monotonically decreases from the left side to the right side, but the electric current density has a complex structure near the cavity, which is related to the electric current vorticity $\bm{\omega}_e = \omega_e \hat{z}$ shown by Fig.~\ref{fig:FEM}~(e).
The electric current vorticity $\omega_e$ is plotted after converted into the commonly-defined vorticity $\omega = \omega_e / e n_{\mathrm{Pt}}$ (in the frequency unit), where $n_{\mathrm{Pt}} = 1.6 \times 10^{28}~\mathrm{/m^3}$ is the electron density of pure $\mathrm{Pt}$~\cite{fischer1980} (we assumed that the electron density of $\mathrm{Pt}_{0.6} (\mathrm{MgO})_{0.4}$ is not so different from that of pure $\mathrm{Pt}$).
The inset of Fig.~\ref{fig:FEM}~(e) is the close-up profile of the vorticity near the upper cavity.
Since the radius of the cavity in the spin Hall system is set as $r = 50~\mathrm{nm}$, which is much larger than the spin diffusion length $\lambda_{\mathrm{s}} = 1.75~\mathrm{nm}$, the sufficient vorticity is only obtained in the vicinity of the cavity structure~[Fig.~\ref{fig:FEM}~(f)].
The vorticity in the upper and lower cavity takes the positive and negative values, respectively, which is comprehended as the spin accumulation due to the direct spin Hall effect in the conventional viewpoint.

Figure~\ref{fig:skyrmion} shows the snapshots of the micromagnetic simulation result, which indicates the successful creation of a magnetic skyrmion.
We compute the Landau-Lifshitz-Gilbert equation numerically by using the FEM result of the vorticity as the input, where we have presumed that the conduction electron spin in the spin Hall system is coupled to the magnetization in CMI through the $sd$-type exchange interaction, $H_{sd} = J_{sd} \bm{s} \cdot \bm{m}$, where $\bm{m}$ is the unit vector of the magnetization in CMI, $\bm{s} = \mathscr{D} \mu_{\mathrm{s}}^z \hat{z}$ is the spin density with the density of state $\mathscr{D}$, and $J_{sd}$ is the strength of the interaction.
A candidate material for the CMI is $\mathrm{Cu}_2 \mathrm{O} \mathrm{SeO}_3$~\cite{langner2014}, and we use the $\mathscr{D}_{\mathrm{Pt}} \sim 3.3 \times 10^{28}~\mathrm{/ eV \, m^3}$~\cite{papaconstantopoulos2015} of pure $\mathrm{Pt}$ as $\mathscr{D}$.
The detailed description of the micromagnetic simulation is given in SI~\cite{SM}.

We find that the magnetic skyrmion is successfully created, and the skyrmion creation is accomplished by the effective magnetic field due to the spin accumulation near the boundary of the cavity through the $sd$-type exchange interaction.
The locally-induced strong effective field creates the magnetic skyrmion~\cite{koshibae2015}.
This mechanism of skyrmion creation is essentially different from the conventional mechanisms based on the spin-transfer torque~\cite{iwasaki2013,iwasaki2013a,sampaio2013}, and the spin-orbit torque~\cite{buttner2017}.
Note that we apply the external magnetic field perpendicular to the layers (the $z$-direction) to stabilize the magnetic skyrmion, but that is not essential for the skyrmion creation.
We also note that the previous works on the skyrmion creation is based on the spin-vorticity coupling~\cite{fujimoto2021,fujimoto2022}, which originates from the general relativity, and the present work is based on the spin-orbit coupling that comes from the special relativity.
It may be interesting to discuss the current vortex generation~\cite{lange2021,fujimoto2021a} from the hydrodynamic viewpoint.

We comment on the previous works on the connection between spin and vorticity~\cite{matsuo2013,takahashi2016,kobayashi2017,matsuo2017,tatara2018,tabaeikazerooni2020,takahashi2020,tabaeikazerooni2021,tatara2021}.
Based on the general relativistic theory, it was shown that the electron spin couples to the mechanical rotation, or vorticity in the local picture, through the spin-rotation (spin-vorticity) coupling~\cite{hehl1990,matsuo2013,matsuo2017}, which was experimentally observed in liquid metals~\cite{takahashi2016,tabaeikazerooni2020,takahashi2020,tabaeikazerooni2021} and by surface acoustic waves~\cite{kobayashi2017,kurimune2020}.
An attempt to interpret the spin-orbit coupling as the spin-vorticity coupling was made based on microscopic calculations~\cite{tatara2018,tatara2021}.

In summary, we have shown that the electron transport in electron systems giving the spin Hall effect is described by the Stokes equation as in electron hydrodynamics.
The kinetic viscosity is determined by the spin diffusion length and the transport lifetime.
The electric current vorticity is proportional to the spin accumulation due to the spin Hall effect in the 2D spin Hall system, while the vorticity in the three-dimensional systems partially captures the spin accumulation.
We discuss the boundary condition between the spin Hall system and the vacuum, which differs from the conventional electron hydrodynamics originating from the electron-electron interaction.
To clarify the difference, we consider the typical example of 2D pipe flow and demonstrate the correspondence between the vorticity and the spin accumulation.
Finally, we show the magnetic skyrmion creation in the bilayer system, which consists of the spin Hall system and the chiral magnetic insulator with cavity structure.
We hope that our theory and demonstration will stimulate both researchers in electron hydrodynamics and spintronics.

This work is supported by CREST Grant Nos. JPMJCR19J4, JPMJCR1874, JPMJCR20C1, and JPMJCR20T1 from JST, Japan, and by JSPS KAKENHI Grant Nos. JP20K03810, JP20H01865, and JP22K13997 from MEXT, Japan.

\bibliography{soc-hydro}
\end{document}